\documentclass[amsmath,amssymb,aps,prc,twocolumn,showpacs,superscriptaddress]{revtex4}
\usepackage[dvips]{graphicx}
\usepackage{times}
\usepackage{color}
\usepackage{amssymb}

\newcommand{\expv}[1]{\langle #1 \rangle}

\newcommand{\diff}[2]{\frac{d #1}{d #2}}

\newcommand{\sbrace}[1]{\left( #1 \right)}
\newcommand{\mbrace}[1]{\left\{ #1 \right\}}
\newcommand{\bbrace}[1]{\left[ #1 \right]}

\newcommand{\Slash}[1]{\ooalign{\hfil/\hfil\crcr$#1$}}
\usepackage{booktabs}

\newcommand{\beq}{\begin{eqnarray}}
\newcommand{\eeq}{\end{eqnarray}}
\newcommand{\psibar}{\overline{\psi}}
\newcommand{\btau}{\boldsymbol{\tau}}
\newcommand{\bpi}{\boldsymbol{\pi}}
\newcommand{\brho}{\boldsymbol{\rho}}
\newcommand{\Lag}{{\cal L}}

\newcommand{\rhoB}{\rho_{\scriptscriptstyle B}}
\newcommand{\rhoBt}{\rho_{{\scriptscriptstyle B}}^\alpha}
\newcommand{\rhoBn}{\rho_{ {\scriptscriptstyle B}}^n}
\newcommand{\rhoBp}{\rho_{ {\scriptscriptstyle B}}^p}
\newcommand{\rhoS}{\rho_{\scriptscriptstyle S}}

\newcommand{\rhoSt}{\rho_{{\scriptscriptstyle S}}^\alpha}
\newcommand{\rhoSn}{\rho_{ {\scriptscriptstyle S}}^n}
\newcommand{\rhoSp}{\rho_{ {\scriptscriptstyle S}}^p}
\newcommand{\rhoT}{\rho_\tau}

\def\nuc#1#2{\relax\ifmmode{}^{#1}{\hbox{#2}}\else${}^{#1}$#2\fi}
\newcommand{\comments}[1]{}
\newcommand{\fpi}{f_\pi}
\newcommand{\mn}{M_{\scriptscriptstyle N}}
\begin{document}
\title {Nuclear Matter and Finite Nuclei in the Effective Chiral Model 
}
\author {P. K. Sahu{\footnote {email:  
pradip@iopb.res.in}}}
\affiliation{
Institute of Physics, 
Sachivalaya Marg, Bhubaneswar 751 005, India}
\affiliation{
Yukawa Institute for Theoretical Physics, Kyoto University,
Kyoto 606-8502, Japan} 
\author{K. Tsubakihara}
\affiliation{
Department of Physics, Faculty of Science,
Hokkaido University, Sapporo 060-0810, Japan}
\author{A. Ohnishi}
\affiliation{
Yukawa Institute for Theoretical Physics, Kyoto University,
Kyoto 606-8502, Japan} 


\begin{abstract}
We systematically investigate the vacuum stability and nuclear properties
in the effective chiral model with higher order terms in $\sigma$.
We evaluate the model parameters by considering the saturation properties
of nuclear matter as well as the normal vacuum to be globally stable
at zero and finite baryon densities.
We can find parameter sets giving moderate equations of state,
and apply these models to finite nuclei.
\end{abstract}
\maketitle

\section{Introduction}

Relativistic mean field (RMF) theory is 
a powerful approach in describing the properties
of infinite nuclear matter and finite nuclei
simultaneously~\cite{SW86,NL1,NL1a,NL3,TM1,Glueball,Glueball1,Glueball2,Glueball3,Glueball4,Schramm02,TO_2007,TO_2007a,Tsubakihara09,Boguta1983,sahu93,sahu00,SO,SJPP,JM}.
While RMF is very successful 
at around the saturation density and in finite nuclei
within the range of the model parameters,
there are uncertainties in the stiffness and the high density behavior
of the equation of state (EOS).
Stiffness of EOS can be 
experimentally probed in heavy-ion collisions~\cite{HIC,HIC1,HIC2,HIC3,HIC4}
and the giant monopole resonances~\cite{GMR,GMR1},
and is crucial in predicting compact astrophysical phenomena.
Therefore, it is very important to develop relativistic models
with constraints to this uncertainty by some symmetries,
and to investigate bulk properties of nuclear systems.
One of the most important symmetries in hadron physics is the chiral symmetry.
Chiral symmetry is a fundamental symmetry in QCD with massless quarks,
and its spontaneous breaking generates hadron masses
through the chiral condensate $\langle q\bar{q} \rangle$~\cite{NJL,NJL1,NJL2},
which is considered to be partially restored in dense matter.
Thus, theories of quarks and hadrons should respect this essential symmetry
in inquiring dense hadronic matter,
and several attempts have been made by including
chiral symmetry in the relativistic nuclear many-body
theories~\cite{Glueball,Glueball1,Glueball2,Glueball3,Glueball4,Schramm02,TO_2007,TO_2007a,Tsubakihara09,Boguta1983,sahu93,sahu00,SO,SJPP,JM}.

The first RMF model was proposed to deal with 
the properties of nuclear matter and finite nuclei~\cite{SW86}.
In this approach, the meson fields (sigma and omega) are treated as classical
limit termed as mean field approximation.
The RMF model is extended by introducing 
the isovector-vector meson $\rho$
and the non-linear self coupling terms of mesons
to obtain better descriptions of nuclear matter 
and finite nuclei~\cite{NL1,NL1a,NL3,TM1}.
The significant point of this extension
is that its non-linear terms simulate the three body forces
which is essential to reproduce the nuclear matter saturation properties
in non-relativistic calculations.
There are several problems in these RMF models
to be regarded as finite density hadronic field theories.
First, these models do not possess chiral symmetry.
While the in-medium nucleon mass is shifted by the $\sigma$ field
and the scalar $\sigma$ field may be related to the chiral condensate,
the self-energy of $\sigma$ is not chirally invariant.
Secondly, the vacuum is not stable with respect to the variation of 
$\sigma$ in many of the RMF models.
While we do not have practical problems in describing nuclear matter
and finite nuclei
in the mean field treatment
where the fluctuation of the meson fields are omitted,
the vacuum instability is conceptually problematic.

The effective chiral model is analogous to the RMF model.
We start from the $\phi^4$ theory, in which the spontaneous symmetry breaking
is included, and vector mesons are introduced in order to describe
the repulsive potential at high density.
When we na\"ively introduce the vector meson field into the $\phi^4$ theory,
however, it is known that the normal vacuum jumps
to a chiral restored abnormal vacuum (Lee-Wick vacuum)
below the saturation density~\cite{LeeWick_1974,Boguta1983},
and this problem is referred to as
the chiral collapse problem~\cite{Thomas2004}.
One of the prescriptions to avoid this problem is to incorporate
a logarithmic term of $\sigma$~\cite{Glueball,Glueball1,Glueball2,Glueball3,Glueball4,Schramm02,TO_2007,TO_2007a,Tsubakihara09}
in the chiral potential
(energy density as a function of $\sigma$ at zero baryon density).
The logarithmic $\sigma$ potential is first introduced
to simulate the scale anomaly in QCD,
which is represented by the glueball dilaton field ($\chi$)
which couples with the logarithm of $\sigma$ as $\propto -\chi^4\log\sigma$%
~\cite{Glueball,Glueball1,Glueball2,Glueball3,Glueball4,Schramm02}.
It is also possible to derive it from the strong coupling limit (SCL)
of lattice QCD~\cite{TO_2007,TO_2007a,Tsubakihara09,SCL,SCL1}.
The logarithmic potential term, $\propto -\log\sigma$, generally
prevents the normal vacuum from collapsing and hence it has no instabilities.
Also this model can describe well the even-even finite nuclei and infinite
nuclear matter properties by including the vector mesons,
their linear couplings to nucleons, and a self-interaction term,
$(\omega_\mu\omega^\mu)^2$~\cite{Schramm02,TO_2007,TO_2007a,Tsubakihara09}.
%
In Ref.~\cite{Schramm02}, for example,
Schramm applied a chiral $\mathrm{SU}(3)$ RMF model
with a logarithmic $\sigma$ potential and the glueball interaction
to nuclei over the whole known range assuming axial symmetry,
and it was demonstrated that
binding energies of spherical and deformed nuclei are well explained
with the precision of $0.1-1$ \% accuracy.
%
While these models have met phenomenological successes
and somewhat based on QCD,
one may doubt the validity of the logarithmic potential,
since it is divergent when the chiral symmetry is restored, $\sigma \to 0$.

Another way to avoid the chiral collapse is to introduce
a dynamical generation of the isoscalar-vector meson mass
through the coupling between scalar and vector
mesons~\cite{Boguta1983,sahu93}.
Since the vector meson becomes light when the chiral symmetry is partially
restored, repulsive effects from the vector meson become strong
and we can avoid the chiral collapse.
One of the problems in this theoretical treatment
is the unrealistically high incompressibility value,
$K \gtrsim 700~\mathrm{MeV}$.
In order to make moderate value of incompressibility,
Sahu {\em et al.} introduced the higher order terms of scalar meson,
$\sigma^6$ and $\sigma^8$~\cite{sahu00,SO,SJPP}.
In this way, we can reproduce the empirical values of
the saturation density, binding energy, and incompressibility
in symmetric nuclear matter.
The advantage of higher order terms in the chiral potential is that
we have a freedom to adopt weaker repulsive vector force
in the nuclear interaction and therefore,
one can have a choice of desirable values of incompressibility.
In the earlier works~\cite{sahu00,SO,SJPP,JM},
the above model with higher order terms in scalar-field interactions
was extensively used in dense matter as well as hot nuclear matter.
In all these works, the vacuum stability at large $\sigma$ values
was not critically examined for all sets of parameters.
It has been recently pointed out~\cite{TO_2007,TO_2007a}
that one of the parameters in the previous work~\cite{SO}
has instability at large $\sigma$.
Though several sets of parameters of the above model were tabulated,
we find that few of them overcome the instability at large $\sigma$.
Therefore, it motivates us to revisit this model and put the stringent 
constrains on parameters for stability with respect to sigma field.

In this paper, we systematically investigate the vacuum stability
and nuclear properties in the effective SU(2) chiral model
having higher order terms in the chiral potential $V_\sigma$.
The condition of the vacuum stability is elucidated
in the parameter plane in the model.
We calculate the EOS for symmetric nuclear matter
with moderate choice of the incompressibility between 300 and 400 MeV,
the effective masses around 0.85 of nucleon mass,
under the constraint of vacuum stability.
The parameters are chosen accordingly by constraining the above
conditions at nuclear saturation points.
We also apply our model to finite spherical nuclei,
and perform a naive dimensional analysis (NDA) to examine the
naturalness of the effective Lagrangian.

All through the paper, we work in the chiral limit ($m_\pi=0$) for simplicity.
As long as we do not explicitly include $\pi$ meson effects
in the mean field approximation,
the results with finite $m_\pi$ in nuclear matter and finite nuclei
are found to be very similar to those in the chiral limit.
We ignore these small differences,
since the main aim of this paper is to elucidate the vacuum stability
condition in effective chiral models.

The paper is organized as follows:
In Sec.~\ref{Sec:Model}, we present a brief formalism 
of the effective chiral model.
We investigate the properties of nuclear matter and finite nuclei
in Sec.~\ref{Sec:Results}.
Basically, we determine the suitable parameters
of the model to explain the saturation properties of nuclear matter
as well the vacuum stability.
Then we determine the equation of state and the properties of finite nuclei.
We also examine the naturalness of the effective Lagrangian
using naive dimensional analysis in Sec.~\ref{Sec:Results}.
We summarize our results in Sec.~\ref{Sec:Summary}.

\section{The Formalism of SU(2) Effective Chiral Model}
\label{Sec:Model}

The effective chiral Lagrangian which includes 
a dynamically generated mass of the isoscalar-vector field, 
$\omega_{\mu}$, that couples to the conserved baryonic current
$j_{\mu}=\bar{\psi}\gamma_{\mu}\psi$ can be written 
as \cite{sahu00,SO},
\begin{align}
\label{lag}
{\cal L} =& \bar{\psi} \left[
	i\Slash\partial
        -g_\sigma(\sigma + i\gamma_5 \btau\cdot{\bpi})
 	-g_\omega\Slash\omega
	-g_\rho{\Slash\brho}\cdot\btau
        \right] \psi \nonumber \\
+&\frac{1}{2}\big(
         \partial_{\mu} \bpi \cdot \partial^{\mu} \bpi
        +\partial_{\mu} \sigma \partial^{\mu} \sigma
        \big)
	- V_\sigma(\sigma,\bpi)
\nonumber \\
-& \frac{1}{4} F^{\mu\nu} F_{\mu\nu}
+ \frac{1}{2}g_{\sigma\omega}^2 x^2 \omega_{\mu}\omega^{\mu}
\nonumber \\
-&\frac{1}{4}{\bf G}_{\mu\nu}\cdot{\bf G}^{\mu\nu}
+\frac{1}{2}m^2_{\rho}{\brho}_{\mu}\cdot{\brho}^{\mu}
\ .
\end{align}
We introduce a chiral symmetric type interaction up to eighth order of the meson
field which reads,
\begin{align}
V_\sigma
=& \frac{C_4\fpi^4}{4}\left(\frac{x^2}{\fpi^2}-1\right)^2
+  \frac{m_\pi^2}{2}x^2 - m_\pi^2\fpi\sigma
\nonumber\\
+&  \frac{C_6\fpi^4}{6}\left(\frac{x^2}{\fpi^2}-1\right)^3
+  \frac{C_8\fpi^4}{8}\left(\frac{x^2}{\fpi^2}-1\right)^4
\ ,
\label{Eq:Vsigma}
\end{align}
where $x^2=\sigma^2+\bpi^2$,
$\fpi$ is the pion decay constant,
$F_{\mu\nu}\equiv\partial_{\mu}\omega_{\nu}-\partial_{\nu}\omega_{\mu}$
and 
${\bf G}_{\mu\nu} \equiv \partial_{\mu}{\brho}_{\nu}-\partial_{\nu}
{\brho}_{\mu}+g_{\rho\rho}\brho_\mu \times \brho_\nu$
are the field tensors of isoscalar- and isovector-vector mesons
($\omega$ and $\rho$-mesons),
and $\psi$, $\bpi$, $\sigma$ denote 
the nucleon isospin doublet, isovector-pseudoscalar pion,
and the scalar fields, respectively.
Coupling constants of nucleon with scalar and vector fields
are introduced as $g_{\sigma}$, $g_{\omega}$ and $g_\rho$, respectively.
We work in natural units where $\hbar = c = k_{\scriptscriptstyle B} = 1$.

The interaction terms of the nucleon and vector meson with
the scalar and pseudoscalar mesons generate the masses
of the nucleon and vector meson
through the spontaneous breaking of the chiral symmetry.
The masses of the nucleon and vector meson in vacuum are given by,
\begin{eqnarray}
\mn = g_{\sigma} \fpi\ ,~~
m_{\omega} = g_{\sigma\omega} \fpi\ ,
\label{Eq:masses}
\end{eqnarray}
where the vacuum expectation value of the $\sigma$ field is replaced
with $\fpi$.
The coefficient $C_4$ is related to the vacuum mass of $\sigma$ as
\begin{align}
C_4 = \frac{m_\sigma^2-m_\pi^2}{2\fpi^2}
\ .
\end{align}
The constant parameters $C_6$ and $C_8$ are included in the 
higher-order self-interaction of the scalar field to describe 
the desirable values of nuclear matter properties at saturation point.
In this work,
we consider the chiral limit,
where the pion mass $m_\pi$ is zero.
In the mean-field treatment we ignore the explicit role of $\pi$ mesons.

By adopting mean-field approximation, the equation of motion of fields 
are obtained from the chiral Lagrangian.
This approach has been used extensively to evaluate the 
EOS~\cite{HIC,HIC1,HIC2,HIC3,HIC4} in many of the theoretical models
for high density matter.
Using the mean-field ansatz in uniform matter,
the equations of motion for the vector fields ($\omega$ and $\rho$-mesons)
are solved as,
\begin{align}
\omega=&\frac{g_\omega^2 \rhoB^2}{g_{\sigma\omega}^2x^2}
=\frac{g_\omega^2 \rhoB^2}{m_\omega^2Y^2}
\ ,
\quad
R\equiv\rho^3_0=\frac{g_\rho}{m_\rho^2} (\rho_p-\rho_n)\ ,
\end{align}
where $Y=\sigma/\fpi$ is the reduction ratio of the chiral condensate
from its vacuum value.
Proton and neutron densities ($\rho_p$ and $\rho_n$) are given as,
$\rho_\alpha = \gamma[k_F^{(\alpha)}]^3/6\pi^2$,
where $k_F^{(\alpha)}$ is the Fermi momentum of the proton
($\alpha=p$) or the neutron ($\alpha=n$),
and $\gamma$ is the spin degeneracy factor, $\gamma=2$.
The baryon density is the sum of proton and neutron density,
$\rhoB=\rho_p+\rho_n$.

The EOS is calculated from the diagonal components of the
conserved total energy-momentum tensor corresponding to the Lagrangian
together with the mean-field equation of motion for the fermion field and a
mean-field approximation for the meson fields.
The total energy density ($\varepsilon$) and pressure ($P$) of
the uniform many-nucleon system are given by,
\begin{align}
\label{ep0}
\varepsilon
=&
	\varepsilon_N(\mn^\star)
	+V_\sigma
	+\frac{g_\omega^2\rhoB^2}{2m_\omega^2Y^2}
	+\frac12m_\rho^2 R^2
\ ,\\
\label{ept}
P =&
	P_N(\mn^\star)
	-V_\sigma
	+\frac{g_\omega^2\rhoB^2}{2m_\omega^2Y^2}
	+\frac12m_\rho^2 R^2
\ ,\\
\varepsilon_N
        =& \sum_{\alpha=p,n} \frac{\gamma}{2\pi^2}
                \int _0^{k_F^{(\alpha)}} k^2dk\sqrt{{k}^2 + \mn^{\star 2}}
\ ,\\
P_N
        =& \sum_\alpha \frac{\gamma }{6\pi^2}
                \int_0^{k_F^{(\alpha)}} \frac{k^4dk}{\sqrt{{k}^2 + \mn^{\star 2}}}
\ ,
\end{align}
where $\mn^{\star} \equiv Y\mn = g_\sigma\sigma$
is the effective mass of the nucleon.
The free (kinetic) nucleon energy density and pressure,
$\varepsilon_N$ and $P_N$,
depend on the effective mass $\mn^\star$,
as well as on the nuclear density $\rhoB$.

The equilibrium value of the scalar field ($\sigma$)
is obtained from the equation of motion,
and it is equivalent to the minimum energy density condition,
$\partial\varepsilon/\partial\sigma=0$.
The equation of motion in terms of $Y$ is given as,
\begin{align}
C_4(1-Y^2)-C_6(1-Y^2)^2+C_8(1-Y^2)^3
\nonumber\\
+\frac{g_\omega^2\rhoB^2}{m_\omega^2\fpi^4Y^4}
-\frac{g_\sigma\rhoS}{\fpi^3Y}=0\ ,
\label{effmass}
\end{align}
where $\rhoS$ denotes the scalar density, defined as
\begin{equation}
\rhoS= \sum_{\alpha=p,n}\frac{\gamma}{(2\pi)^3}
	\int^{k_F^{(\alpha)}}_0
	\frac{\mn^\star d^3k} {\sqrt{k^2+{\mn^\star}^2}}
\ .
\end{equation}

In the previous works~\cite{sahu00,SO,SJPP,JM},
the $\omega N$ coupling and $\sigma\omega$
coupling was assumed to be the same, $g_\omega=g_{\sigma\omega}$,
and the pion decay constant $\fpi$ was not introduced explicitly.
The energy density is represented by
$c_\sigma(=g_\sigma^2/m_\sigma^2), c_\omega=g_\omega^2/m_\omega^2$,
and $B$ and $C$, which are related to the $\sigma^6$ and $\sigma^8$
coefficients, respectively.
In these works, implicitly given $\fpi$ value is not necessarily
the same as the observed one, $\fpi \simeq 93~\mathrm{MeV}$.
However, it is possible to map those parameters into the coefficients
in the present work by comparing, for example, the energy density,
if we do not require the condition, $g_\omega=g_{\sigma\omega}$.
The relation with their parameters with the present parameters
are given by,
\begin{align}
C_4=\frac{\mn^2}{2c_\sigma \fpi^4} \ ,\ 
C_6=\frac{B}{2\fpi^4c_\sigma c_\omega} \ ,\ 
C_8=\frac{C}{2\fpi^4c_\sigma c_\omega^2\mn^2} \ ,
\end{align}
and $g_\omega = \sqrt{c_\omega} m_\omega$.
In the later discussion, we also examine the stability of their Lagrangians.

\section{Results and discussions}
\label{Sec:Results}

In this section we will determine the parameter sets at 
nuclear matter saturation density.
We will select the parameters
by examining the stability with respect to sigma field.
We will then use stable parameter sets to find the properties of
finite spherical nuclei and then examine the naturalness
of the effective Lagrangian by performing the naive dimensional analysis.
Here, we use constants $\mn=938$ MeV, $f_{\pi}=93$ MeV, 
$m_{\omega}=783$ MeV, $m_{\rho}=770$ MeV and $g_{\sigma}=\mn/f_{\pi}$.

\subsection{Fixing parameters in symmetric nuclear matter}

In the effective chiral model,
the chiral symmetry relates the interaction parameters
and hadron masses, and reduces the number of parameters,
as shown in Eq.~(\ref{Eq:masses}).
In the present treatment, we have five parameters,
$g_\omega, g_\rho, C_4, C_6$ and $C_8$.
Here we determine three parameters,
the nucleon coupling to the vector field, $g_\omega$,
and the coefficients in the scalar potential terms, $C_4$ and $C_6$,
in symmetric nuclear matter.
These parameters are obtained as functions of $C_8$ and
the nucleon effective (Landau) mass $\mn^\star(\rho_0)$,
by fitting the empirical saturation point, $(\rho_0, E_0/A)$,
where $E_0/A$ is the binding energy per nucleon at saturation density,
$\rhoB=\rho_0$.
The saturation point plays a decisive role on finite nuclear binding energies
and radii, thus we will adjust them to reproduce finite nuclear property.
Moreover, the incompressibility $K$ and the nucleon effective mass $\mn^\star(\rho_0)$
are the keys in EOS around the saturation point,
as well as at high densities.
The nuclear incompressibility is somewhat uncertain at saturation point.
The desirable values of effective mass and nuclear matter incompressibility
are chosen in accordance with recent heavy-ion collision data~\cite{HIC,HIC1,HIC2,HIC3,HIC4}.
In our calculation,
we have examined several parameter sets corresponding to each incompressibility
and effective mass in the range of
$200-400$ MeV and $(0.8-0.9)M_N$~\cite{moll88}, respectively,
to observe the sensitivity of EOS in the high density region.

\begin{figure}[tb]
\begin{center}
\includegraphics[width=8cm]{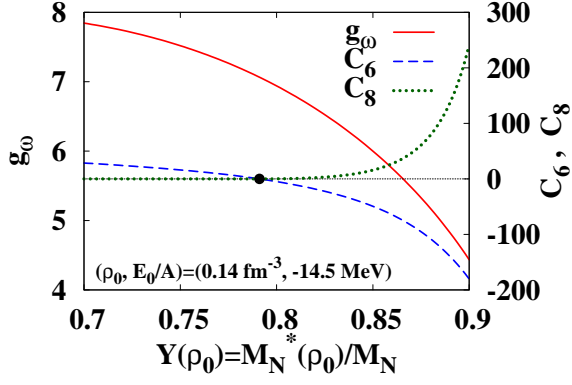}
\end{center}
\caption{(Color online) 
$g_\omega$ as a function of $Y(\rho_0)\equiv\mn^\star(\rho_0)/\mn$ (solid line),
and $C_6$ (dashed line), $C_8$ (dotted line) values on the vacuum stability boundary.
}\label{Fig:gomg}
\end{figure}

\begin{figure}[tb]
\begin{center}
\includegraphics[width=8cm]{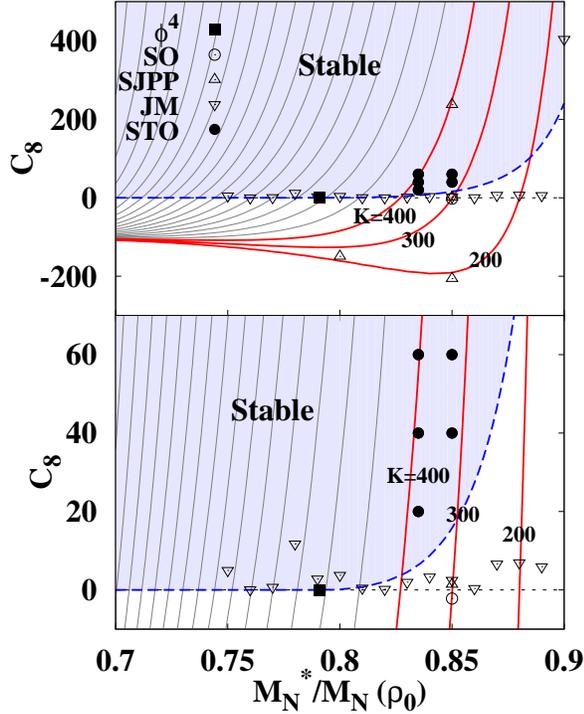}
\end{center}
\caption{(Color online) 
Vacuum stability region (shaded area)
and the incompressibility $K$
in the $(Y(\rho_0)=\mn^\star(\rho_0)/\mn,C_8)$ plane.
We have displayed several other models~\protect{\cite{SO,SJPP,JM}}
(open symbols) with $\phi^4$ theory (filled square)
and present results (filled circles).
}\label{Fig:stab}
\end{figure}

First, we introduce and examine the vacuum stability of chiral potential
as the constraint on the parameters.
Vacuum stability condition of the present effective chiral models can be examined as follows.
The vacuum energy density is given in $V_\sigma$ in Eq.~(\ref{Eq:Vsigma}),
which is rewritten as
\begin{align}
\frac{V_\sigma}{\fpi^4}=&\frac{X^2}{2} f(X)
\,,\ 
f(X)=\frac{C_8}{4} X^2 + \frac{C_6}{3} X + \frac{C_4}{2}
\,,
\end{align}
where $X=(Y^2-1)$.
In stable cases, $V_\sigma$ must be always positive in the range $X>-1$
except for the vacuum $X=0$ (i.e. $\sigma=\fpi$), at which $V_\sigma=0$.
Provided that $C_8 \geq 0$,
the stability is ensured when one of the following conditions is satisfied;
(1) the discriminant of $f(X)$, $D=(C_6/3)^2-4(C_8/4)(C_4/2)$, is negative,
(2) the discriminant of $f(X)$ is positive,
 but the solution of $f(X)=0$ are in the range $X\leq-1$,
or 
(3) in the case of $C_8=0$, $C_6\geq0$ and $f(-1)>0$.

\begin{table*}[tb]
\caption{Effective chiral model parameter sets.
Stability of the model is also shown;
the vacuum is stable against the variation of $\sigma$ in models with "S",
and the vacuum is unstable at large $\sigma$ in models with "U".
}
\label{Table:Pars}
\centering
\begin{tabular}{lccccccccc}
\hline
\hline
	& $\mn^\star/\mn$ &$g_\omega$ &$C_4$ &$C_6$	& $C_8$		& $K$		& $m_\sigma$	& $g_\rho$ & Stability\\
		& 	& 	&	&		&		& (MeV)		& (MeV)		&	& \\
\hline
$\phi^4$	& 0.781 & 6.781 & 37.16 & 0		& 0		& 695.4		& 801.7		& ---	& S\\
STO-1		& 0.850 & 6.001 & 35.32 & -39.24    & 40.00     & 318.5     & 781.7     & 3.597 & S\\
STO-2       & 0.835 & 6.331 & 35.37 & -25.47    & 20.00     & 376.0     & 782.2     & 3.467 & S\\
STO-3       & 0.835 & 6.328 & 36.18 & -16.77    & 40.00     & 389.5     & 791.1     & 3.467 & S\\
STO-4       & 0.835 & 6.328 & 37.10 & -7.682    & 60.00     & 402.3     & 801.1     & 3.467 & S\\
STO-5       & 0.850 & 6.001 & 36.09 & -30.91    & 60.00     & 327.2     & 790.1     & 3.467 & S\\ 
\hline
SO		& 0.85	& 5.610	& 33.60 & -74.380	& -2.200	& 335.300	& 762.3	      	&       & U\\
\hline
SJPP2003-I	& 0.85	& 5.598	& 25.84 & -159.000	& -206.200	& 210.000	& 668.6	      	&       & U\\
SJPP2003-II	& 0.85	& 5.598	& 33.73 & -73.210	& 1.405		& 300.000	& 763.8	      	&       & U\\
SJPP2003-III	& 0.85	& 5.598	& 42.72 & 24.260	& 237.400	& 380.000	& 859.7	      	&       & S\\
SJPP2003-IV	& 0.80	& 6.532	& 26.94 & -92.050	& -149.400	& 300.000	& 682.6	      	&       & U\\
SJPP2003-V	& 0.90	& 4.047	& 98.28 & 906.200	& 4270.000	& 300.000	& 1304.0      	&       & S\\
\hline
JM2009-1	& 0.75	& 7.106	& 38.71 & 17.030	& 4.998		& 1142.000	& 818.3	      	& 4.39 &  S\\
JM2009-2	& 0.76	& 7.016	& 37.87 & 9.958		& 0.085		& 1010.000	& 809.3	      	& 4.40 &  S\\
JM2009-3	& 0.77	& 6.908	& 37.63 & 6.021		& 0.713		& 897.200	& 806.8	      	& 4.41 &  S\\
JM2009-4	& 0.78	& 6.796	& 38.13 & 7.567		& 11.750	& 815.000	& 812.2	      	& 4.42 &  S\\
JM2009-5	& 0.79	& 6.669	& 37.10 & -3.428	& 2.817		& 710.400	& 801.1	      	& 4.43 &  S\\
JM2009-6	& 0.80	& 6.531	& 36.80 & -9.658	& 3.751		& 630.700	& 797.8	      	& 4.44 &  S\\
JM2009-7	& 0.81	& 6.380	& 36.20 & -19.340	& 0.482		& 555.500	& 791.4	      	& 4.45 &  U\\
JM2009-8	& 0.82	& 6.212	& 35.75 & -29.030	& 0.179		& 490.800	& 786.4	      	& 4.46 &  U\\
JM2009-9	& 0.83	& 6.048	& 35.37 & -38.570	& 1.966		& 439.300	& 782.2	      	& 4.47 &  U\\
JM2009-10	& 0.84	& 5.830	& 34.71 & -53.690	& 3.313		& 383.800	& 774.8	      	& 4.48 &  U\\
JM2009-11	& 0.85	& 5.605	& 33.81 & -72.440	& 2.481		& 335.800	& 764.8	      	& 4.49 &  U\\
JM2009-12	& 0.86	& 5.358	& 32.61 & -96.850	& 0.383		& 292.200	& 751.1	      	& 4.49 &  U\\
JM2009-13	& 0.87	& 5.09	& 31.26 & -125.400	& 6.621		& 254.400	& 735.4	      	& 4.50 &  U\\
JM2009-14	& 0.88	& 4.780	& 29.11 & -166.800	& 6.942		& 217.900	& 709.7	      	& 4.51 &  U\\
JM2009-15	& 0.89	& 4.435	& 26.05 & -224.500	& 5.910		& 183.500	& 671.2	      	& 4.52 &  U\\
JM2009-16	& 0.90	& 4.049	& 28.83 & -191.300	& 404.300	& 173.400	& 706.2	      	& 4.53 &  S\\
\hline
\end{tabular}
\end{table*}

%
By fitting the saturation point,
we can fix $C_4$, $C_6$ and $g_\omega$ as functions of
$C_8\geq 0$ and the effective mass at normal density,
$Y(\rho_0) = \mn^\star/\mn \simeq 0.8-0.9$.
First, we give $Y(\rho_0)$,
then $g_\omega$ is uniquely determined.
From Eqs.~(\ref{ep0}) and (\ref{ept}),
we find that the enthalpy density is free from $V_\sigma$,
\begin{align}
\varepsilon + P = &
	\frac{g_\omega^2\rhoB^2}{m_\omega^2Y^2}
	+m_\rho^2 R^2
	+\varepsilon_N(\mn^\star) + P_N(\mn^\star)
\\
	= &\rhoB(\mn - B/A) \quad (\rhoB = \rho_0, \rho_p=\rho_n)\ .
\end{align}
In the second line, we have used the fact that $P=0$
at the saturation density.
This equation only depends on one parameter $g_\omega$,
and we can fix it from the saturation property.
In Fig.~\ref{Fig:gomg}, 
we show $g_\omega$ value as a function of $Y(\rho_0)=\mn^\star(\rho_0)/\mn$.
The energy gain from the $\sigma$ meson is small for larger values of $Y(\rho_0)$,
then the repulsive potential is also chosen to be small
to reproduce the binding energy $B/A$ at $\rho_0$.
Thus $g_\omega$ is a decreasing function of $Y(\rho_0)$.
Next, we give the value of $C_8 \geq 0$.
For a given set of $(Y(\rho_0), C_8)$,
we can solve the condition,
$\varepsilon(\rho_0)=\rho_0(\mn-B/A)$
and $\partial\varepsilon/\partial\sigma=0$ at $\rhoB=\rho_0$
(Eq.~(\ref{effmass})),
with respect to $C_4$ and $C_6$.

From these coefficients, we examine the vacuum stability condition.
In Fig.~\ref{Fig:gomg}, 
we show $C_6$ and $C_8$ values on the vacuum stability boundary
as a function of $Y(\rho_0)$.
These values are equivalent to the minimum $C_6$ and $C_8$ values
for each $Y(\rho_0)$.
At $Y(\rho_0)=0.781=Y_{\phi^4}$, $\phi^4$ theory is realized, i.e. $C_8=C_6=0$.
For larger value of $Y$ at $\rhoB=\rho_0$,
repulsion from $\omega$ is chosen to be smaller as shown in Fig.~\ref{Fig:gomg},
and larger repulsion in $V_\sigma$ is required.
This repulsion can be generated by negative $C_6$ or positive $C_8$,
since $(Y^2-1)^3$ and $(Y^2-1)^4$ are negative and positive
for $Y<1$, respectively.
Thus for small values of $C_8$, negative $C_6$ values are required
at $Y(\rho_0)>Y_{\phi^4}$,
and to keep the vacuum stability, there exists the minimum value of $C_8$.

In Fig.~\ref{Fig:stab}, we show the vacuum stability region
(shaded area) in the $(Y(\rho_0), C_8)$ plane.
We have examined the stability of the parameter sets
proposed in previous works~\cite{SO,SJPP,JM}.
We show these parameter sets
by open symbols in Fig.~\ref{Fig:stab} and in Table \ref{Table:Pars}.
Unfortunately, most of the parameter sets which gives medium
incompressibility ($200~\mathrm{MeV} < K < 400~\mathrm{MeV}$)
are unstable in vacuum against the variation of $\sigma$,
marked with "U" in the last column of Table \ref{Table:Pars}.
Only one parameter set (SJPP-V) fulfills the vacuum stability condition
and gives a medium $K$ value.

Parameter sets which we propose and examine in this paper,
STO-$i$ ($i=1, 2, \cdots 5$),
are tabulated in Table~\ref{Table:Pars},
and shown in filled circles in Fig.~\ref{Fig:stab}.
We have chosen two values of $\mn^\star/\mn$ ($0.835$ and $0.85$),
and three values of $C_8$ (20, 40 and 60).
The combination $(\mn^\star/\mn,C_8)=(0.85, 20)$ is close
to the stability boundary, and we do not adopt them.
All of these parameters give stable chiral potentials,
and the incompressibility are in the range,
$200~\mathrm{MeV} < K \lesssim 400~\mathrm{MeV}$.
We adopt the saturation point
$(\rho_0,E_0/A)=(0.14~\mathrm{fm}^{-3}, -(14.5 \sim 14.6)~\mathrm{MeV})$,
which is found to explain the binding energies of heavy nuclei
reasonably well as discussed in the next subsection.

\begin{figure}[tb]
\begin{center}
\includegraphics[width=8cm]{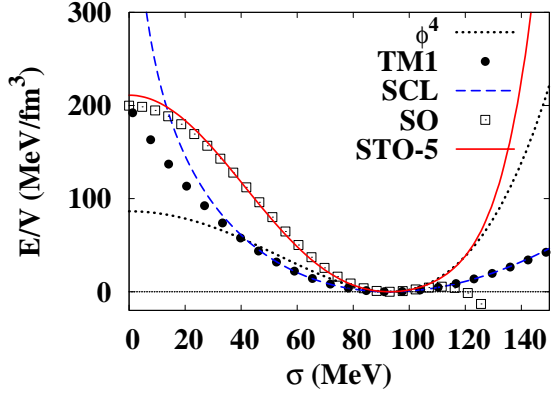}
\end{center}
\caption{(Color online) The chiral potential as a function of $\sigma$
in the $\phi^4$ theory (dotted line),
SCL (dashed line)~\protect{\cite{TO_2007,TO_2007a}},
SO (open square)~\protect{\cite{SO}},
and STO-5 (solid line).
Results of a non-chiral model, TM1 (filled circles)~\protect{\cite{TM1}}
are also shown for comparison.
}\label{Fig:vac}
\end{figure}

\begin{figure}[tb]
\begin{center}
\includegraphics[width=8cm]{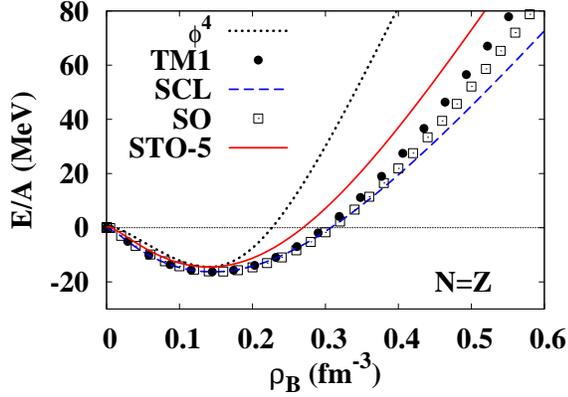}
\end{center}
\caption{(Color online) Energy per nucleon as a function of the baryon density.
Meaning of the lines and symbols are the same as 
in Fig.~\protect{\ref{Fig:vac}}.
}\label{Fig:eos}
\end{figure}

In Fig.~\ref{Fig:vac}, we show the chiral potential $V_\sigma$,
in STO-5 as an example.
We also show the chiral potential
in the $\phi^4$ theory, TM1~\cite{TM1},
SCL~\cite{TO_2007,TO_2007a} and SO~\cite{SO},
for comparison.
The chiral potential in STO-5 behaves similarly to that in SO
in the region $\sigma < \fpi$.
At larger $\sigma$ values, unstable parameters give negative chiral potentials
in the region $\sigma > \fpi$ as shown in the SO case.
In Fig.~\ref{Fig:eos}, we show the EOS in STO-5
in comparison with other EOSs.
We find that EOS in STO-5 is reasonably soft,
and comparable to those in TM1 and SCL,
which explains the bulk properties of finite nuclei.

\subsection{Finite nuclei}

In describing finite nuclei,
it is numerically preferable to represent the Lagrangian in the shifted field
$\varphi \equiv f_\pi - \sigma$
and to separate the $\sigma$ mass term from the chiral potential $V_\sigma$,
since the boundary condition is given as $\varphi \to 0$ at $r\to\infty$.
In addition, it is necessary to include the photon field
which represents the Coulomb potential.
Here we take the static and mean-field approximation for boson fields,
then RMF Lagrangian can be written as follows,
\begin{align}
\Lag_\chi^{\scriptscriptstyle \mathrm{RMF}}=
&\psibar\left[
	i\Slash{\partial} - M_N^\ast(\varphi) - \gamma^0 U_v(\omega,R,A)
  \right]\psi \nonumber\\
&- \frac{1}{2}\left({\bigtriangledown}\varphi\right)^2
                    - \frac{1}{2}m_\sigma^2 \varphi^2 - V_\varphi(\varphi)
\nonumber\\
&+ \frac{1}{2}\left({\bigtriangledown}\omega\right)^2 + \frac{1}{2}M_\omega^2(\varphi) \omega^2
 \nonumber\\
&+ \frac{1}{2}\left({\bigtriangledown} R\right)^2 + \frac{1}{2}m_\rho^2 R^2
 + \frac{1}{2}\left({\bigtriangledown} A\right)^2\ ,\label{RMFlag-f}
\end{align}
where
\begin{gather}
M_N^\ast(\varphi) = M_N-g_{\sigma}\varphi \ ,\ \ 
M_\omega^2(\varphi) = g_{\sigma\omega}^2\left(f_\pi {-} \varphi\right){{}^2}\ ,
\\
U_v(\omega,R,A)= g_{\omega}{\omega} + g_{\rho}\tau_3 R + e\frac{1+\tau_3}{2} A 
\ ,\\
V_\varphi \equiv V_\sigma - \frac12 m_\sigma^2\varphi^2
\ .
\end{gather}
The field equations of motion derived from this Lagrangian read,
\begin{gather}
\bbrace{-i\bold{\alpha}\cdot\bold{\bigtriangledown} + \beta M^\ast + U_v}\psi
        = \varepsilon_i \psi
\ ,\label{Lag-baryon}\\
\sbrace{-\triangle + m_\sigma^2}\varphi
=  g_\sigma\rhoS - \diff{V_\varphi}{\varphi}
  {-} g_{\sigma\omega}^2 \left(f_\pi {-} \varphi\right)\omega^2
\ ,\label{Lag-sigma}\\
\sbrace{-\triangle + m_\omega^2}\omega
= g_\omega\rhoB
  {+} g_{\sigma\omega}^2\varphi\left(2f_\pi {-} \varphi\right)\omega
\ ,\label{Lag-omega}\\
\sbrace{-\triangle + m_\rho^2}R = g_\rho\rhoT
\ ,\label{Lag-rho}\\
-\triangle A =  e\rhoBp
\ ,\label{Lag-photon}
\end{gather}
where $\rhoS=\rhoSp+\rhoSn$, $\rhoB=\rhoBp+\rhoBn$, $\rhoT=\rhoBp-\rhoBn$
denote scalar, baryon and isospin densities of nucleons, respectively.
Total energy is given by the integral of the energy density given as,
\begin{align}
E =&   \sum_{i\kappa\alpha}n^{\mathrm{occ}}_{i\kappa\alpha} \varepsilon_{i\kappa\alpha} \nonumber\\
   & - \frac{1}{2}\int \mbrace{ {-}g_{\sigma}\varphi\rhoS + g_{\omega}\omega\rhoB
     + g_{\rho}R\rhoT + e^2A\rhoBp}d\bold{r}\nonumber\\
   & + \int\mbrace{V_\varphi
     - \frac12\varphi\diff{V_\varphi}{\varphi}
     {-} \frac{g_{\sigma\omega}^2}{2}\varphi\sbrace{f_\pi {-} \varphi}\omega^2
     } d\bold{r}
\label{totalBE}
\end{align}
where we use the Eq.(\ref{Lag-baryon})-(\ref{Lag-photon})
to calculate second order derivatives of meson fields.
Nucleon single particle states are specified by
the radial quantum number $i$, isospin $\alpha(=p, n)$,
and the angular momentum quantum number, 
$\kappa = l$ ($\kappa=-(l + 1)$) for $j = l-1/2$ ($j = l+1/2$).
The number of occupied nucleon is represented by $n^\mathrm{occ}_{i\kappa\alpha}$,
which is equal to $2|\kappa|=2j+1$ for filled single particle states.
We solve the self-consistent coupled equations
(\ref{Lag-baryon})-(\ref{Lag-photon})
by iteration until the convergence of total energy is achieved.
In this work, we assume that the nuclei under consideration are spherical,
then the nucleon wave functions are expanded
in spherical harmonic basis as follows,
\begin{gather}
        \psi_{\alpha i\kappa m}= \sbrace{
        \begin{array}{c}
                i[G^\alpha_{i\kappa}/r]\Phi_{\kappa m}\\
                -[F^\alpha_{i\kappa}/r]\Phi_{-\kappa m}
        \end{array}
        }
        \zeta_\alpha
\ ,\\
        \rhoBt =
            \sum_{i\kappa}\sbrace{\frac{n^\mathrm{occ}_{i\kappa\alpha}}{4\pi r^2}}
                                \sbrace{|G_{i\kappa}^\alpha (r)|^2 + |F_{i\kappa}^\alpha (r)|^2}
\ ,\\
        \rhoSt =
            \sum_{i\kappa}\sbrace{\frac{n^\mathrm{occ}_{i\kappa\alpha}}{4\pi r^2}}
                                \sbrace{|G_{i\kappa}^\alpha (r)|^2 - |F_{i\kappa}^\alpha (r)|^2}
\ ,
\end{gather}
where $\zeta_\alpha$ represents the isospin wave function, $\alpha=p, n$.

In comparing the calculated results in mean-field models
with the experimental binding energies and charge radii, 
we have to take account of several corrections.
In this work, we consider the center-of-mass (CM) kinetic energy correction
on the total energy,
and CM and nucleon size correction
on nuclear charge rms radius in the same way as that adopted in Ref.~\cite{TM1}.
The CM kinetic energy is assumed to be
\begin{equation}
        E_\mathrm{ZPE}
        = \frac{\expv{\bold{P}_\mathrm{CM}^2}}{2AM_N}
	\simeq \frac{3}{4}\hbar\omega
        = \frac{3}{4} 41A^{- 1/3}
\ ,
\label{come}
\end{equation}
where $\bold{P}_\mathrm{CM}=\sum_i \bold{p}_i$ is the CM momentum.
This correction gives an exact result
when the state is represented by a harmonic-oscillator wave function,
and we assume that it also applies to the RMF wave functions.
The CM correction on the proton rms radius 
is written as
\begin{align}
\delta \expv{r_\mathrm{p}^2}
&=- 2\expv{\bold{R}_\mathrm{CM}\cdot\bold{R}_\mathrm{p}}
  +\expv{\bold{R}_\mathrm{CM}^2}
\nonumber\\
&\simeq	\begin{cases}
	{\displaystyle - {3\hbar\over2AM_N\omega}}
	 	&
         \mbox{(for heavy nuclei)\ ,}\\
         {\displaystyle
		   - \frac{2\expv{r_\mathrm{p}^2}}{A}
		   + \frac{\expv{r_\mathrm{m}^2}}{A}
         }
		& \mbox{(for light nuclei)\ ,}
	\end{cases}
\label{comc}
\end{align}
where $\bold{R}_\mathrm{p}=\sum_{i\in p} \bold{r}_i/Z$ is
the proton CM position,
and $\expv{r_\mathrm{p}^2}$ and $\expv{r_\mathrm{m}^2}$ represent
the proton and matter mean square radii, respectively.
We assume again that harmonic-oscillator results applies for heavy nuclei.
For light nuclei, we evaluate the correction
in RMF wave functions, and we consider only the direct-term contributions.
The charge rms radius is obtained
by including the finite size effects of protons and neutrons,
\begin{equation}
        \expv{r_\mathrm{ch}^2}
        = \expv{r_\mathrm{p}^2}
		+ \expv{r_\mathrm{size}^2}_\mathrm{p}
		- {N\over Z} \expv{r_\mathrm{size}^2}_\mathrm{n}
\ ,
        \label{crms}
\end{equation}
where $\expv{r_{\mathrm{size}}^2}_\alpha$ denotes the size of
proton or neutron
and is equal to $(0.862\mathrm{fm})^2$ and $(0.336\mathrm{fm})^2$, respectively.
We evaluate the binding energies and charge rms radii with these corrections,
and the pairing energy for open-shell nuclei are neglected.

In describing finite nuclei, the isospin dependent interaction
is an important ingredient.
In uniform nuclear matter,
we can obtain the symmetric energy coefficient $a_\mathrm{sym}$
by expanding the energy density around the symmetric nuclear matter,
\begin{equation}
a_{\rm sym} = \frac{g_\rho^2 k_F^3}{3\pi^2m_\rho^2}
+ \frac{k_F^2}{6\sqrt{k_F^2+\mn^{\star 2}}}\ ,
\end{equation}
where $k_F=(6\pi^2\rhoB/\gamma)^{1/3} (\rhoB=\rho_p+\rho_n, \gamma=4)$
is the Fermi momentum in symmetric nuclear matter.
In order to reproduce the empirical value of the symmetry energy coefficient,
$a_{\rm sym} = 32 \pm 6~\mathrm{MeV}$~\cite{moll88},
the above relation gives $g_{\rho}=4.625$ for and $M^\star=0.85M$.
In the present work, we have fixed $g_\rho$
by fitting the binding energies of heavy nuclei.
We find that $g_\rho \simeq 3.5$ is appropriate.

\begin{table}[tb]
\caption{Calculated results for saturation property of symmetric nuclear matter,
         $B/A$, charge rms radii of stable nuclei.}
\label{Table:BE}
\begin{tabular}{c|ccccc|c}
\hline
\hline
\multicolumn{7}{l}{Saturation property}\\
\hline
               & STO-1   & STO-2      & STO-3      & STO-4      & STO-5      & Exp.  \\
\hline
$\rho_0$       &  0.14  &  0.14     &  0.14     &  0.14     &  0.14     &       \\
$E_0$          & -14.6  & -14.5     & -14.6     & -14.6     & -14.5     &       \\
K              &  318.5 &  376.0    &  389.5    &  402.3    &  327.2    &       \\
\hline
\multicolumn{7}{l}{$B/A$}\\
\hline
$^{12 }$C      & 9.38   & 9.40      & 9.37      & 9.26      & 9.30      & 7.68  \\
$^{16 }$O      & 10.9   & 10.9      & 10.9      & 10.7      & 10.8      & 7.98  \\
$^{28 }$Si     & 9.53   & 9.57      & 9.87      & 9.47      & 9.47      & 8.45  \\
$^{40 }$Ca     & 9.91   & 9.85      & 9.56      & 9.80      & 9.86      & 8.55  \\
$^{48 }$Ca     & 9.73   & 9.74      & 9.74      & 9.68      & 9.70      & 8.67  \\
$^{90 }$Zr     & 9.13   & 9.11      & 9.15      & 9.09      & 9.11      & 8.71  \\
$^{116}$Sn     & 8.83   & 8.81      & 8.85      & 8.80      & 8.81      & 8.52  \\
$^{196}$Pb     & 7.87   & 7.85      & 7.91      & 7.86      & 7.86      & 7.87  \\
$^{208}$Pb     & 7.87   & 7.85      & 7.91      & 7.86      & 7.87      & 7.87  \\
\hline
\multicolumn{7}{l}{Charge rms radii}\\
\hline
$^{12 }$C      & 2.27   & 2.27      & 2.27      & 2.27      & 2.28      & 2.46  \\
$^{16 }$O      & 2.44   & 2.44      & 2.44      & 2.44      & 2.44      & 2.74  \\
$^{28 }$Si     & 2.95   & 2.94      & 2.95      & 2.95      & 2.95      & 3.09  \\
$^{40 }$Ca     & 3.30   & 3.30      & 3.30      & 3.30      & 3.30      & 3.45  \\
$^{48 }$Ca     & 3.39   & 3.39      & 3.39      & 3.39      & 3.39      & 3.45  \\
$^{90 }$Zr     & 4.21   & 4.20      & 4.20      & 4.20      & 4.21      & 4.26  \\
$^{116}$Sn     & 4.59   & 4.59      & 4.59      & 4.59      & 4.60      & 4.63  \\
$^{196}$Pb     & 5.48   & 5.47      & 5.47      & 5.47      & 5.48      &  -    \\
$^{208}$Pb     & 5.56   & 5.55      & 5.54      & 5.54      & 5.56      & 5.50  \\
\hline                                                                  
\end{tabular}
\end{table}

We have tuned the model parameters so as to explain
the binding energies of heavy nuclei.
In Table \ref{Table:BE}, we tabulate finite nuclei results.
We also summarize the STO parameter sets
and the resulting incompressibility in Table \ref{Table:Pars}.
The binding energy per nucleon in heavy nuclei
and the incompressibility parameters would be in the acceptable range.
On the contrary,
we cannot reproduce the binding energies of light nuclei
simultaneously with heavy nuclei.
Calculated values of charge rms radii underestimate the data
as long as we vary the $\rho_0$ value in the acceptable range,
$\rho_0=0.14-0.16~\mathrm{fm}^{-3}$.

\begin{figure}[tb]
\begin{center}
\includegraphics[width=8cm]{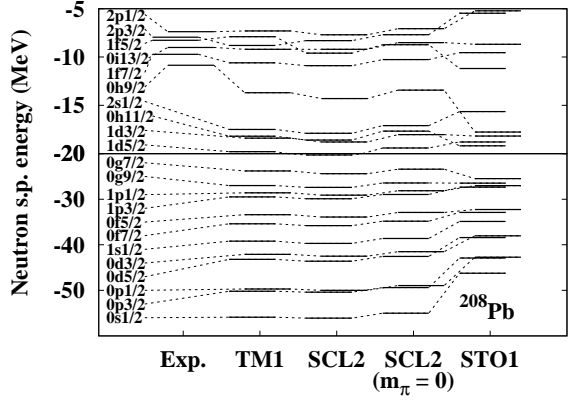}
\end{center}
\caption{Neutron single particle levels of ${}^{208}\mathrm{Pb}$
in TM1, SCL2, SCL2 in the chiral limit, and STO models.}
\label{Fig:spl}
\end{figure}

While we show the results obtained in the chiral limit ($m_\pi=0$),
the results are not modified much with finite pion mass
in the mean field approximation.
This is examined in 
an RMF model with the chiral SU$(2)$ logarithmic potential (SCL2)
with finite $m_\pi$~\cite{TO_2007}.
In Fig.~\ref{Fig:spl}, we present calculated neutron single particle levels
of $^{208}$Pb in the chiral limit.
Not only in the level structure but also in their energies,
there is little differences between
the SCL2 results with finite and zero pion masses.
%
Therefore in the mean field approximation,
we can safely discuss the properties of the RMF Lagrangian
in the chiral limit.

We also find in Fig.~\ref{Fig:spl} that
the present model (STO-1) show smaller 
$ls$ splitting compared with other RMF models
due to large effective nucleon masses.
One of the original motivations to use RMF models for nuclei was
the large $ls$ splitting naturally generated
from the large scalar and vector potentials additively.
In the recently developed chiral RMF models~\cite{Schramm02,TO_2007},
$ls$ splittings are evaluated to be smaller than empirical values.
This may be suggesting the need to include explicit pion effects.
There are some discussions regarding the contribution to
the $ls$ splitting~\cite{KW}
where one-pion exchange tensor force and two-pion exchange with the excitation
of virtual $\Delta(1232)$ isobars are taken into account for explaining
$ls$ splitting property of nucleon and hyperon simultaneously.
The $ls$-like roles of the tensor force or pions
are discussed also in light nuclei~\cite{TensorLS}.
We have not taken account of these pion effects in our present calculations,
then $\pi$ treatment may be required in order
to resolve the $ls$ splitting problem in chiral RMF models
and effective chiral models.

\subsection{Naive dimensional analysis}
\label{Subsec:NDA}

The present STO model is a kind of effective field theory,
contains higher order terms and is non-renormalizable.
Then it would be valuable to examine the naturalness
in the naive dimensional analysis (NDA)\cite{Furnstahl1996,Saito1997b,MG84,MG84a,MG84b}.
It is found that
the loop contributions with the momentum cutoff $\Lambda \sim 1\ \mathrm{GeV}$
generate the following terms
with dimensionless coefficients $C_{lmnp}$ of order unity,
\cite{MG84,MG84a,MG84b,Furnstahl1996}
\begin{align}
  \mathcal{L}_\mathrm{int}
  \sim& \sum_{l,m,n,p}\frac{C_{lmnp}}{m!n!p!}
                   \sbrace{\frac{\bar\psi\Gamma\psi}{f_\pi^2\Lambda}}^l
\nonumber\\
	&\times
                   \sbrace{\frac{\varphi}{f_\pi}}^m
                   \sbrace{\frac{\omega}{f_\pi}}^n
                  \sbrace{\frac{\rho  }{f_\pi}}^p
                   (f_\pi\Lambda)^2\ ,
		   \label{Eq:NDA}
\end{align}
where $\Gamma$ denotes the $\gamma$ and $\tau/2$ when necessary.

An effective theory having terms in Eq.~(\ref{Eq:NDA})
is considered to hold naturalness,
when all the dimensionless coefficients $C_{lmnp}$ are of order unity.
In the present effective Lagrangian,
we obtain the following dimensionless coefficients,


\begin{align}
&C_{1100} = \frac{f_\pi g_\sigma}{\Lambda} = \frac{\mn}{\Lambda}
	\sim 0.94\ ,\nonumber \\ 
&C_{1010} = \frac{f_\pi g_\omega}{\Lambda}
	\sim 0.56\ , \nonumber \\ 
&C_{1001} = \frac{2f_\pi g_\rho}{\Lambda}
	\sim 0.64\ ,\nonumber \\ 
&C_{0120} = - \frac{2g_{\sigma\omega}^2\fpi^2}{\Lambda^2}
	= - \frac{2m_\omega^2}{\Lambda^2}
	\sim 1.2\ ,\nonumber \\ 
&C_{0220} = \frac{2g_{\sigma\omega}^2\fpi^2}{\Lambda^2}
	= \frac{2m_\omega^2}{\Lambda^2}
	\sim 1.2\ ,\nonumber \\ 
&C_{0300} = \frac{\fpi^2}{\Lambda^2}\,3!\,\left(\frac43 C_6 - C_4\right)
	\sim -4
\ ,\nonumber\\
&C_{0400} = \frac{\fpi^2}{\Lambda^2}\,4!\,\left(
		2C_8-2C_6+\frac14 C_4
	\right) \sim 40
\ ,\nonumber\\
&C_{0500} = \frac{\fpi^2}{\Lambda^2}\,5!\,\left(
		-4C_8+C_6
	\right) \sim -280
\ ,\nonumber\\
&C_{0600} = \frac{\fpi^2}{\Lambda^2}\,6!\,\left(
		3C_8-\frac16C_6
	\right) \sim  1200
\ ,\nonumber\\
&C_{0700} = - \frac{\fpi^2}{\Lambda^2}\,7! C_8
	\sim -2600
\ ,\nonumber\\
&C_{0800} = \frac{\fpi^2}{\Lambda^2}\,\frac{8!}{8} C_8
	\sim 2600
\ .\nonumber\\
\end{align}
We show the results in STO-5, and adopt $\Lambda=1~\mathrm{GeV}$.
We find that the meson-nucleon and $\sigma\omega$ couplings are natural,
but the self-interaction coefficients in $\sigma$ are not natural.

\section{Summary and conclusions}
\label{Sec:Summary}

In this paper, we have investigated the properties of nuclear matter
and finite nuclei in the effective chiral model
with $\sigma^6$ and $\sigma^8$ terms.
The nucleon-vector meson coupling is found to be uniquely determined
as a function of the effective mass at normal nuclear matter density,
$Y(\rho_0)\equiv\mn^\star(\rho_0)/\mn$,
and we have specified the region of stability
in the $(Y(\rho_0), C_8)$ plane, where $C_8$ is the coefficient of
the $\sigma^8$ term.
We can find the parameter sets which satisfies the vacuum stability
condition and results in moderate incompressibility,
$K=(200-400)~\mathrm{MeV}$.
The incompressibility is found to be dominated
by the nucleon effective mass,
and $\mn^\star(\rho_0)/\mn \gtrsim 0.83$ is necessary
in order to obtain moderate $K$,
as far as the vacuum stability is required.

The obtained effective chiral model with higher order terms in $\sigma$
is applied to finite nuclei for the first time.
We can explain the binding energies of heavy nuclei (Sn and Pb)
reasonably well, while we overestimate the binding energies of light nuclei.
This may be because the nucleon-vector meson coupling is small,
$g_{\omega N} \sim 6$ compared with other RMF models
which explains nuclear binding energies in a wide mass range,
such as
NL1 ($g_{\omega N}=13.285$)~\cite{NL1,NL1a},
NL3 ($g_{\omega N}=12.868$)~\cite{NL3},
TM1 ($g_{\omega N}=12.6139$)~\cite{TM1},
and
SCL ($g_{\omega N}=13.02$)~\cite{SCL,SCL1}.
Smaller $g_{\omega N}$ value is compensated at higher densities
where the chiral symmetry is partially restored and $\omega$ mass decreases,
but light nuclei are more sensitive to the EOS at lower densities.

We have also performed the na\"ive dimensional analysis 
(NDA)~\cite{Furnstahl1996,Saito1997b,MG84,MG84a,MG84b}
of the present model.
Moderate $K$ value of around 300 MeV requires the $\sigma^8$
coefficient $C_8 \gtrsim 20$ as found in Fig.~\ref{Fig:stab}.
This value corresponds to $C_{0800}\gtrsim 870$,
and the model cannot hold naturalness.
In order to construct effective chiral models
having moderate incompressibility, vacuum stability and naturalness
simultaneously, it would be necessary to introduce other types of
interaction terms other than polynomial forms of $\sigma$.
Works in this direction would be valuable for the understanding
of the chiral properties of the QCD vacuum.

\section*{Acknowledgment}
 
This work was supported in part by
the Grant-in-Aid for Scientific Research
from MEXT and JSPS
under the grant numbers,
   17070002,	
   19540252,	
and
   20-4326,	
the Yukawa International Program for Quark-hadron Sciences (YIPQS),
and the Global COE Program "The Next Generation of Physics, Spun from 
Universality and Emergence". KT also thanks JSPS for the fellowship (20-4326).

\end{document}